\documentclass[11pt,bookmarks=false]{article}

\usepackage[colorlinks,citecolor=blue,urlcolor=blue]{hyperref}
\usepackage{amssymb}
\usepackage{amsthm}
\usepackage{amsmath}
\usepackage{booktabs}
\usepackage{mathrsfs}
\usepackage{multirow}
\usepackage{bm}
\usepackage{geometry}
\usepackage{graphics}
\usepackage{float}
\usepackage{epsfig}
\usepackage{subfigure}
\usepackage{enumerate}

\numberwithin{equation}{section}
\newtheorem{thm}{Theorem}[section]

\newtheorem{prop}[thm]{Proposition}

\def\proclaim#1{\par \bigskip\noindent {\bf #1}\bgroup\it\ }
\def\endproclaim{\egroup\par\bigskip}

\newcommand{\dif}{\mathrm{d}}
\def\pr{\mathbb{P}}

\def\text#1{\mbox{\rm #1}}

\begin{document}
\title{\bf A Normal Approximation Method for  Statistics in Knockouts}
\date{}
\maketitle
\begin{center}
\vskip -1.5cm {{\sc {\large Yutong Nie}\footnote[1]{Department of Mathematics, Zhejiang University, Hangzhou 310007, P.R.China; ytnie@zju.edu.cn}  and  {\large Chenhe Zhang}\footnote[2]{Department of Mathematics, Zhejiang University, Hangzhou 310007, P.R.China; chhzhang@zju.edu.cn}}}
\end{center}

\bigskip
\noindent{\bf Abstract.} 
The authors give an approximation method for Bayesian inference in arena model, which is focused on paired comparisons with eliminations and bifurcations. The approximation method simplifies the inference by reducing parameters and introducing normal distribution functions into the computation of posterior distribution, which is largely based on an important property of normal random variables. Maximum a posteriori probability (MAP) and Bayesian prediction are then used to mine the information from the past pairwise comparison data, such as an individual's strength or volatility and his possible future results. We conduct a simulation to show the accuracy and stability of the approximation method and demonstrate the algorithm on nonlinear parameter inference as well as prediction problem arising in the FIFA World Cup.


\section{Introduction}
Pairwise comparisons play an essential role in the real world and has pervaded into all areas of life. For example, physiological reactions, match results in sports, preference between items, and species competitions are all caused by pairwise comparisons in a certain sense. In the last century, a great deal of effort in statistical modeling was devoted to the study of pairwise comparisons. In 1927, Thurstone \cite{PA} first studied a psychological continuum where two physical stimulus magnitudes are compared. Two decades later, it became a significant topic in sports; Bradley and Terry \cite{RAI} proposed a probability model to predict the outcomes of paired comparisons and Elo \cite{RCP} developed a rating system to update ranks of players. After that, there has been extensive study and application of pairwise comparisons, such as dynamic Bradley-Terry models concerning changeable merits \cite{DBTM,DSM,DPC,PEL} and algorithms for ranking \cite{RCDS,MMA}. 

However, in recent years, some new questions and critics also emerged. It is naural to ask  ``how much the outcome of a match is influenced by skill, or by chance'', as presented in \cite{MCS}. Besides, as Aldous stated in \cite{ERS}, ``there has been surprisingly little `applied probability' style
mathematical treatment of the basic model''. To solve some of these problems, Zhang introduced an original parametric model in \cite{myself} called arena model, which essentially is a type of latent variable model. Arena provides a framework of statistics in knockouts, such as FIFA World Cup, which is mainly concerned with the estimation of an individual's strength and quantification of volatility. However, it only studies two simplest arenas: $m$-$n$ arena without fluctuations and 1-1 arena with uniform fluctuations. In this paper, we present an estimation method for the general case, that is $m$-$n$ arena with fluctuations. Due to the complicated expression of the likelihood when considering fluctuations for large $m$ or $n$, we conduct Bayesian inference based on some approximate results to simplify the estimation. Through assuming uniform fluctuations and computing likelihood by normal distribution functions, we obtain a series of results which match the true values surprisingly in simulations.

The rest of the paper is organized as follows. In Section 2, we review some basic concepts and conclusions in arena model. The difficulties of classic methods and reasons for using approximation methods are stated in Section 3. This is followed in Section 4 by the Bayesian inference on samples from an arena with fluctuations. In Section 5, we discuss how to predict individuals' future results from past data along this path. Finally, the estimates and predictions given by the approximation method are evaluated by simulations and applications in the FIFA World Cup.

\section{A quick review of arenas}

The concept of arena is introduced by Zhang (2018) through an ideal game following four basic rules, which are (R1)-(R4) in \cite{myself}. To apply that concept into statistical inference, four general model assumptions are proposed to establish an $m$-$n$ arena without fluctuations.

(A1) In an arena, an infinite number of \emph{runs} can be held among a fixed group of individuals. These individuals are called \emph{players}. All players constitute a countably infinite set $A_{0,0}^{q}=\{a_1,a_2,\cdots\}$, where $a_l$ is the $l$-th player and $q=1,2,\cdots$.

(A2) Each player has an observable \emph{state} $(i,j)\in\varepsilon$ with respect to time and an unobservable constant \emph{strength} $x\in\mathbb{R}$, where
$$\varepsilon=\{(i,j):0\leqslant i\leqslant m,0\leqslant j\leqslant n\}\backslash\{(m,n)\}.$$
Denote the strength of the $l$-th player by $X_l$. Assume $X_1,X_2,\cdots\,X_n,\cdots$ are independent and identically distributed, supported on $\Theta$, and their density function is $p(x)$.

(A3) Let $A_{i,j}^{q}$ denote the set of players whose states are $(i,j)$ after $(i+j)$-th round in the $q$-th run. If
$$0\leqslant i\leqslant m-1,0\leqslant j\leqslant n-1,$$
then the system will \emph{randomly assign} him an opponent $a_{l'}$ from $A_{i,j}^{q}$. If $X_{l}>X_{l'}$, then let
$$a_l\in A_{i+1,j}^{q}, a_{l'}\in A_{i,j+1}^{q}.$$
Otherwise, let
$$a_l\in A_{i,j+1}^{q}, a_{l'}\in A_{i+1,j}^{q}.$$

(A4) If a player's state satisfies $i=m$ or $j=n$ in the $q$-th run, then we say the player's $q$-th run is over and the $(i,j)$ is called his \emph{result} of the $q$-th run. When all players' $q$-th runs are over, a new run will start according to (A3). At the same time, their numbers of runs $q$ plus one.

\begin{figure*}[!htb]
\centering
\subfigure[Elimination form] {\includegraphics[height=2in,width=2.8in,angle=0]{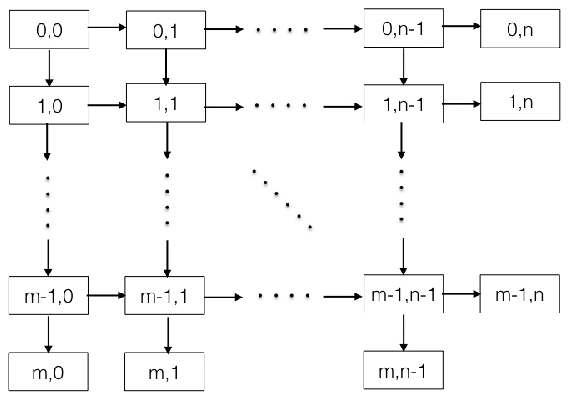}}
\subfigure[Bifurcation form] {\includegraphics[height=2in,width=2.8in,angle=0]{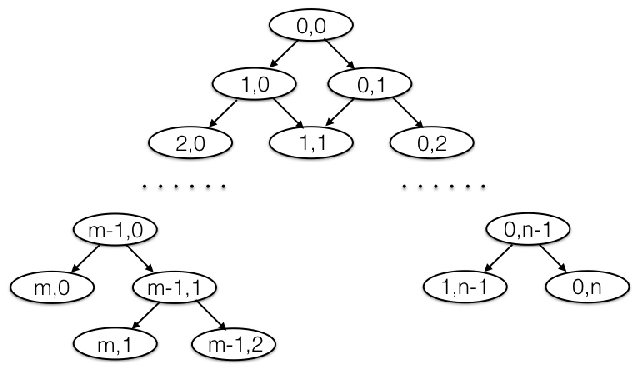}}
\caption{The figures illustrate how players in arena flow dynamically in one arena run.}
\end{figure*}

For $m$-$n$ arenas without fluctuations, \cite{myself} described the strengths of players in the state $(i,j)$ by a random variable $X_{i,j}$, whose probability density is $p_{i,j}(\cdot)$ and derived an invariant Bayesian predictor for future results. To do the inference for a knockout, we first need to compute the probability distribution of $X_{i,j}$, the strength of a player who has reached state $(i,j)$ in a run. The recursion equation is given by

\begin{equation}
\left\{\begin{aligned}
p_{0,0}(x)&=p(x),\\
p_{i,0}(x)&=2p_{i-1,0}(x)\int_{-\infty}^xp_{i-1,0}(t)dt,\qquad(1 \leqslant i \leqslant m-1)\\
p_{0,j}(x)&=2p_{0,j-1}(x)\int_x^{+\infty}p_{0,j-1}(t)dt,\qquad(1 \leqslant j \leqslant n-1)\\
p_{i,j}(x)&=\frac{2i}{i+j}p_{i-1,j}(x)\int_{-\infty}^{x}p_{i-1,j}(t)dt+\frac{2j}{i+j}p_{i,j-1}(x)\int_x^{+\infty}p_{i,j-1}(t)dt,\\ 
&\quad(1\leqslant i\leqslant m-1,1\leqslant j\leqslant n-1)\\
p_{m,j}(x)&=2p_{m-1,j}(x)\int_{-\infty}^xp_{m-1,j}(t)dt,\qquad(0 \leqslant j \leqslant n-1)\\
p_{i,n}(x)&=2p_{i,n-1}(x)\int_x^{+\infty}p_{i,n-1}(t)dt,\qquad(0 \leqslant i \leqslant m-1)
\end{aligned}\right.
\label{e dens of str}
\end{equation}

where $p(\cdot)$ is the density in assumption (A2). It is easy yield the CDF (denoted by $F_{i,j}(\cdot)$) of $X_{i,j}$ from the probability density of $p_{i,j}(\cdot)$, which is directly applied into the Bayesian inference of a player's strength, given his past performance. We have

\begin{equation}
\left\{
\begin{aligned}
F_{0,0}(x)&=\int_{-\infty}^{x}p(t)\dif t,\\
F_{i,0}(x)&=\bigl(F_{i-1,0}(x)\bigr)^2,\qquad(0 \leqslant i \leqslant m-1)\\
F_{0,j}(x)&=1-\bigl(1-F_{0,j-1}(x)\bigr)^2,\qquad(0 \leqslant j \leqslant n-1)\\
F_{i,j}(x)&=\frac{i}{i+j}\bigl(F_{i-1,j}(x)\bigr)^2+\frac{j}{i+j}\bigl(1-\bigl(1-F_{i,j-1}(x)\bigr)^2\bigr),\\ 
&\quad(1 \leqslant i \leqslant m-1,1 \leqslant j \leqslant n-1)\\
F_{m,j}(x)&=\bigl(F_{m-1,j}(x)\bigr)^2,\qquad(0 \leqslant j \leqslant n-1)\\
F_{i,n}(x)&=1-\bigl(1-F_{i-1,n}(x)\bigr)^2.\qquad(0 \leqslant i \leqslant m-1)
\end{aligned}\right.
\label{e dist of str}
\end{equation}

Suppose we have $k$ samples of a player's final results in an $m$-$n$ arena without fluctuations, namely $(i_{1},j_{1}),(i_{2},j_{2}),\cdots,(i_{k},j_{k})$. To compute the likelihood, it requires to know the probability that an individual with strength $x$ obtains different final results. The conditional probability mass function is given by

\begin{equation}
\begin{split}
\mathbb{P}(\mathcal{A}_{m,j}|X=x)&=\binom{m+j-1}{m-1}(\frac{1}{2})^{m+j}\cdot \frac{p_{m,j}(x)}{p_{0,0}(x)},  \qquad j=0,1,\cdots,n-1,\\
\mathbb{P}(\mathcal{A}_{i,n}|X=x)&=\binom{n+i-1}{n-1}(\frac{1}{2})^{n+i}\cdot \frac{p_{i,n}(x) }{p_{0,0}(x)}, \qquad i=0,1,\cdots,m-1.
\end{split}
\label{e prob with infer}
\end{equation}

If one choose $p(x)$ as the prior distribution of the player's strength, then we have
\begin{equation}
\pi(\lambda|\tilde{x})=\frac{\prod_{t=1}^k\mathbb{P}(\mathcal{A}_{i_t,j_t}|X=\lambda)p(\lambda)}{\int_\Theta \prod_{t=1}^k\mathbb{P}(\mathcal{A}_{i_t,j_t}|X=\lambda)p(\lambda)\dif\lambda}
\label{e post dest}
\end{equation}
as the posterior distribution of his strength, given the past several results. Finally, the prediction of his future performance can be done by combining equation (\ref{e prob with infer}) with the posterior distribution we already obtain.

In $m$-$n$ arenas with fluctuations, another parameter called coefficient of fluctuations joins in. And accordingly the assumptions are revised to 

(A1) In an arena, an infinite number of \emph{runs} can be held among a fixed group of individuals, and these individuals are called \emph{players}. All players constitute a countably infinite set $A_{0,0}^{q}=\{a_1,a_2,\cdots\}$, where $a_l$ is the $l$-th player and $q=1,2,\cdots$.

(A2') For each player, there is an observable $(i,j)\in\varepsilon$ as his \emph{state} with respect of time and an unobservable constant $x\in\mathbb{R}$ as his \emph{strength}, where
$$\varepsilon=\{(i,j):0\leqslant i\leqslant m,0\leqslant j\leqslant n\}\backslash\{(m,n)\}.$$
Denote the strength of the $l$-th player as $X_l$. Assume $X_1,X_2,\cdots\,X_n,\cdots $ are independent and identically distributed, and their density function is $p(x)$. Let
\begin{equation}
X_l^{q,k}=X_l+\frac{\rho_l}{\sqrt{2}}\epsilon_l^{q,k}
\end{equation}
be the performance of the $l$-th player in the $k$-th round of his $q$-th run, where $\rho_l>0$ is an unknown value called the coefficient of fluctuations of the $l$-th player and $\epsilon_{l}^{q,k}$ is the relative fluctuations of the $l$-th player in the $k$-th round of the $q$-th run. Assume 
$$\epsilon_l^{1,1},\epsilon_l^{1,2},\cdots,\epsilon_l^{2,1},\epsilon_l^{2,2},\cdots,\epsilon_l^{3,1},\epsilon_l^{3,2},\cdots i.i.d \sim N(0,1)$$
and $X_l,\epsilon_l^{q,k},\epsilon_{l'}^{q',k'}$ are mutually independent for arbitrary $q,q', k,k'$ and $l\neq l'$.

(A3') Let $A_{i,j}^{q}$ denote the set of players whose states are $(i,j)$ after $(i+j)$-th round in the $q$-th run. If
$$0\leqslant i\leqslant m-1,0\leqslant j\leqslant n-1,$$
then the system will \emph{randomly assign} an opponent $a_{l'}$ from $A_{i,j}^{q}$ to him. If $X_{l}^{q,i+j+1}>X_{l'}^{q,i+j+1}$, then let
$$a_l\in A_{i+1,j}^{q}, a_{l'}\in A_{i,j+1}^{q}.$$
Otherwise, let
$$a_l\in A_{i,j+1}^{q}, a_{l'}\in A_{i+1,j}^{q}.$$

(A4') If a player's state satisfies $i=m$ or $j=n$ in the $q$-th run, then we say the player's $q$-th run is over and this state $(i,j)$ is called his \emph{result} of the $q$-th run. When all players' $q$-th runs are over, a new run will start according to (A3'). At the same time, their numbers of runs $q$ plus one.

Actually Zhang \cite{myself} only studies 1-1 arenas with uniform fluctuations and has not discussed the general cases. Assume in a 1-1 arena with uniform fluctuations, $m$ players are sampled randomly and their results $I_{lk}=\mathbf{1}$\{The $l$-th player wins his $k$-th round\} form an $m\times n$ sample matrix
\[I=\left(\begin{array}{cccc}
I_{11}&I_{12}&\cdots&I_{1n}\\
I_{21}&I_{22}&\cdots&I_{2n}\\
\vdots&\vdots&\ddots&\vdots\\
I_{m1}&I_{m2}&\cdots&I_{mn}\\
\end{array}\right).\] then a strongly consistent estimator of the coefficient of fluctuations is given by 
\begin{equation}
\hat{\rho}=\sqrt{\frac{3-tan^2\pi T}{tan^2\pi T-1}},
\label{e esti of infty}
\end{equation}
where $T=\frac{1}{n-1}\left(\frac{1}{mn}\sum_{l=1}^{m}Y_{l}^2-\frac{1}{2}\right)$ and $Y_l=\sum_{k=1}^{n}I_{lk},l=1,2,\cdots,m$. In this paper, we focus on providing an estimation method for general $m$-$n$ arena with fluctuations. Now we present our results step by step.

\section{Difficulties and solutions}

As previously stated, arena model is aimed to infer the strength and coefficient of fluctuations for each individual, according to their past performance in an arena. \cite{myself} only studies the estimation in two simplest arenas: m-n arena without fluctuations and 1-1 arena with uniform fluctuation. In this paper, we will extend the work for the general case of $m$-$n$ arenas with fluctuations. But first let us begin with a natural generalization, that is 1-1 arena with ``ununiform'' fluctuations, where the coefficients of fluctuations of individuals are not necessarily equal. Provided that we already know the final results of $M$ players in $N$ runs in an 1-1 arena with fluctuation. And suppose the strength and coefficient of fluctuations of a randomly chosen individual has a joint CDF $F(x,a)$, then the probability that a randomly chosen player win each run (one's final result in a 1-1 arena can only be (1,0) or (0,1), which are called ``win'' and ``lose'' for simplicity) is given by

\begin{equation}
    \begin{split}
    \mathbb{P}(I_{1k}=1|X_1=x,\rho_1=a)=&\int_{0}^{+\infty}\int_{-\infty}^{+\infty}\mathbb{P}(I_{1k}=1|X_1=x,\rho_1=a,X'=y,\rho'=b)\dif F(y,b)\\
    =&\int_{0}^{+\infty}\int_{-\infty}^{+\infty}\mathbb{P}(x+\frac{a}{\sqrt{2}}\epsilon_{1k}>y+\frac{b}{\sqrt{2}}\epsilon_k')\dif F(y,b)\\
    =&\int_{0}^{+\infty}\int_{-\infty}^{+\infty}\mathbb{P}(\frac{b\epsilon_k'-a\epsilon_{1k}}{\sqrt{a^2+b^2}}<\frac{x-y}{\sqrt{(a^2+b^2)/2}})\dif F(y,b)\\
    =&\int_{0}^{+\infty}\int_{-\infty}^{+\infty}\Phi(\frac{x-y}{\sqrt{(a^2+b^2)/2}})\dif F(y,b).
    \end{split}
    \label{e cond_prob}
\end{equation}

Therefore, an estimate of $\pr(I_{1k}=1)$ can only reflects the complicated relationship between $X_1$ and $\rho_1$ rather than specific estimates of themselves. On the other hand, consider that the result of a player obeys a uniform distribution of win and loss. This is equally likely to be caused by his medium strength and his extremely high coefficient of fluctuations. The feasibility of estimation in 1-1 arena with uniform fluctuations is due to an additional restriction that all individuals' coefficient of fluctuations equal. In a word, inference in 1-1 arena with ``ununiform'' fluctuations, which seems to be an easy work, is not applicable instead. However, it is possible to do such inference for the case that either $m>1$ or $n>1$. Notice that a player with high coefficient of fluctuations tends to gain both good results (such as ``$m$-0'') and bad results (such as ``0-$n$''), while a player with low one performs more steady, even though uncertainty and chaos are also partly resulted from random matching. How can we give a metric to quantify such chaos and fluctuations? We will consider this uncertainty quantification problem in the following parts.

\subsection{Difficulties of ``exact estimation''}

Intuitively, suppose the final results of an individual in $k$ runs in an $m$-$n$ arena with fluctuations ($m\geqslant 2$ or $n\geqslant 2$) are $(i_1,j_1),(i_2,j_2),\cdots,(i_k,j_k)$, then the sample variance of $i_1-j_1,i_2-j_2,\cdots,i_k-j_k$ is a reasonable reflection of his fluctuations. Nevertheless, there lacks a direct connection between the value and the player's coefficient of fluctuations. It could be an approach worth study but we do not follow that way in this paper. Notice that one of our ultimate goals is to predict an individual's future results given his past performance, which requires $\mathbb{P}(\mathcal{A}_{m,j}|X_l=x,\rho_l=a)\ (j=1,2,\cdots,n-1)$ and $\mathbb{P}(\mathcal{A}_{i,n}|X_l=x,\rho_l=a)\ (i=1,2,\cdots,m-1)$. We first derive their expressions here.

\begin{thm}
	In an $m$-$n$ arena with fluctuations, consider a fixed player $a_l$. Suppose his strength and coefficient of fluctuations are respectively $X_l$ and $\rho_l$. Denote the event that the final result of a run in the arena is $(i,j)$ by $\mathcal{A}_{i,j}\ (i=m$ or $j=n)$. Provided that the strengths and coefficients of fluctuations of all players who have reached the state $(i,j)\ (0\leqslant i\leqslant m,0\leqslant j\leqslant n)$ have a joint PDF $p_{i,j}(x,a)$, we have 
	\begin{equation}
	\mathbb{P}(\mathcal{A}_{m,j}|X_l=x,\rho_l=a)=\mathbb{P}(\mathcal{A}_{m,j})\cdot \frac{p_{m,j}(x,a)}{p_{0,0}(x,a)},  \qquad j=0,1,\cdots,n-1,
	\label{e prob with infer21}
	\end{equation}
	\begin{equation}
	\mathbb{P}(\mathcal{A}_{i,n}|X_l=x,\rho_l=a)=\mathbb{P}(\mathcal{A}_{i,n})\cdot \frac{p_{i,n}(x,a)}{p_{0,0}(x,a)}, \qquad i=0,1,\cdots,m-1.
	\label{e prob with infer22}
	\end{equation}
	Here,
	\[\begin{split}
	\mathbb{P}(\mathcal{A}_{m,j})=&\binom{m+j-1}{m-1}(\frac{1}{2})^{m+j}, \qquad j = 0,1,\cdots,n-1,\\
	\mathbb{P}(\mathcal{A}_{i,n})=&\binom{n+i-1}{n-1}(\frac{1}{2})^{n+i}, \qquad i = 0,1,\cdots,m-1.
	\end{split}\]
	\label{t prob with infer2}
\end{thm}

\begin{proof}
The definition of $\mathcal{A}_{i,j}$ and $X_{i,j}$ yield
\begin{equation*}
\begin{split}
\mathbb{P}(\mathcal{A}_{m,j}|X_l=x,\rho_l=a)&=\lim\limits_{(\Delta x,\Delta \rho) \rightarrow (0,0)}\frac{\mathbb{P}(\mathcal{A}_{m,j},x<X\leqslant x+\Delta x, a<\rho_l\leqslant \rho+\Delta \rho)}{\mathbb{P}(x<X\leqslant x+\Delta x, a<\rho\leqslant a+\Delta \rho)}\\
&=\lim\limits_{(\Delta x,\Delta \rho) \rightarrow (0,0)} \frac{\mathbb{P}(x<X_{m,j}\leqslant x+\Delta x, a<\rho_{m,j}\leqslant a+\Delta \rho)\mathbb{P}(\mathcal{A}_{m,j})}{p_{0,0}(x,a)\Delta x \Delta \rho}\\
&=\mathbb{P}(\mathcal{A}_{m,j})\cdot \frac{p_{m,j}(x,a)}{p_{0,0}(x,a)},\qquad j=0,1,\cdots ,n-1.
\end{split}
\end{equation*}
By the same token, we can obtain equation (\ref{e prob with infer22}).
\end{proof}

Notice that all of the $p_{i,j}(x,a)$ are unknown or have not been estimated so far, including $p_{0,0}(x,a)$. Combine with equation (\ref{e cond_prob}) and imitate the proof of Theorem 2.2 in \cite{myself}, we can yield the following recursion equation of $p_{i,j}(x,a)$:

\begin{thm}
	Let $(X_{i,j},\rho_{i,j})$ describe the strength and coefficient of fluctuations of a player who reaches the state $(i,j)$ in a run. Suppose $(X_{0,0},\rho_{0,0})$ is a continuous random vector, then $(X_{i,j},\rho_{i,j})$ are continuous random vectors, and satisfy
	\begin{equation}
	\left\{\begin{aligned}
	p_{i,0}(x,a)=&2p_{i-1,0}(x,a)\int_{-\infty}^{+\infty}\int_{0}^{+\infty}\Phi(\frac{x-y}{\sqrt{(a^2+b^2)/2}})p_{i-1,0}(y,b)\dif y\dif b,\qquad(1 \leqslant i \leqslant m-1)\\
	p_{0,j}(x,a)=&2p_{0,j-1}(x,a)\int_{-\infty}^{+\infty}\int_{0}^{+\infty}\Phi(\frac{x-y}{\sqrt{(a^2+b^2)/2}})p_{0,j-1}(y,b)\dif y\dif b,\qquad(1 \leqslant j \leqslant n-1)\\
	p_{i,j}(x,a)=&\frac{2i}{i+j}p_{i-1,j}(x,a)\int_{-\infty}^{+\infty}\int_{0}^{+\infty}\Phi(\frac{x-y}{\sqrt{(a^2+b^2)/2}})p_{i-1,j}(y,b)\dif y\dif b\\
	&+\frac{2j}{i+j}p_{i,j-1}(x,a)\int_{-\infty}^{+\infty}\int_{0}^{+\infty}\Phi(\frac{x-y}{\sqrt{(a^2+b^2)/2}})p_{i,j-1}(y,b)\dif y\dif b,\\ 
	&(1\leqslant i\leqslant m-1,1\leqslant j\leqslant n-1)\\
	p_{m,j}(x)=&2p_{m-1,j}(x,a)\int_{-\infty}^{+\infty}\int_{0}^{+\infty}\Phi(\frac{x-y}{\sqrt{(a^2+b^2)/2}})p_{m-1,j}(y,b)\dif y\dif b,\qquad(0 \leqslant j \leqslant n-1)\\
	p_{i,n}(x)=&2p_{i,n-1}(x,a)\int_{-\infty}^{+\infty}\int_{0}^{+\infty}\Phi(\frac{x-y}{\sqrt{(a^2+b^2)/2}})p_{i,n-1}(y,b)\dif y\dif b,\qquad(0 \leqslant i \leqslant m-1)
	\end{aligned}\right.
	\label{e dens of str1}
	\end{equation}
	where $p_{i,j}(\cdot,\cdot)$ is the joint PDF of $(X_{i,j},\rho_{i,j})$. 
	\label{t dens of str1}
\end{thm}

The theorem above tells us that we could compute the probability that a player with strength $x$ and coefficient of fluctuations $a$ obtains different final results only if we know $p_{0,0}(x,a)$ for each $x$ and $a$. However, it is impractical to give any analytical expression of $p_{i,j}(x,a)$ even if the distribution of all players' coefficient of fluctuations is degenerate, let alone estimate parameters by this way. Hence, we must give up this theoretically exact but practically ineffective approach and resort to some approximation methods.

Our goal at present is to give an estimate of an individual's coefficient of fluctuations. A natural simplification is to assume an $m$-$n$ arena with uniform fluctuations. It sounds weird but we have sufficient reasons to do this way. First, if the only data we have is the past final results of an individual within several runs, we know little about the information of other players. It is better to reduce an integral by supposing uniform fluctuations. Otherwise we will have to solve expensive computations just by making some seemingly reasonable but still false assumptions. Furthermore, the uniformity assumption at least provides a raw but easy estimate, which does not hurt to be optimized by subsequent iterations. We focus on giving rough estimates of strengths and coefficients of fluctuations in this paper, and leave the optimization study to some further research.

\subsection{A computationally efficient approximation}

After assuming the uniformity of fluctuations, the equation \ref{e dens of str1} reduces to (take the first one as an instance)

\begin{equation}
p_{i,0}(x)=2p_{i-1,0}(x)\int_{-\infty}^{+\infty}\Phi\bigl(\frac{x-y}{\rho}\bigr)p_{i-1,0}(y)\dif y,\qquad(1 \leqslant i \leqslant m-1),
\label{e pi0(x)}
\end{equation}
where, $p_{i,j}(\cdot)$ is the probability density of $X_{i,j}$. Since it involves the convolution of normal $\Phi$ function and probability density, we first prove a property of normal random variables.

\begin{prop}
	Suppose $\xi\sim N(0,1)$, $a,b$ are two constants. Then
	\begin{equation}
	\mathbb{E}\Phi(a+b\xi)=\Phi\bigl(\frac{a}{\sqrt{1+b^2}}\bigr).
	\label{e exphi}
	\end{equation}
	\label{t exphi}
\end{prop} 

\begin{proof}
	Define
	\begin{equation}
	f(a)=\mathbb{E}\Phi(a+b\xi)=\int_{-\infty}^{+\infty}\Phi(a+bx)\frac{1}{\sqrt{2\pi}}e^{-\frac{x^2}{2}}\dif x.
	\label{e defn f}
	\end{equation}
	Since
	$$g(a,x)=\frac{1}{\sqrt{2\pi}}\Phi(a+bx)e^{-\frac{x^2}{2}}\geqslant 0$$
	is continuously differentiable on $\mathbb{R}^2$ and
	$$\int_{-\infty}^{+\infty}g(a,x)\dif x\leqslant \int_{-\infty}^{+\infty}\frac{1}{\sqrt{2\pi}}e^{-\frac{x^2}{2}}\dif x=1<\infty,$$
	$$\int_{-\infty}^{+\infty}\left|\frac{\partial}{\partial a}g(a,x)\right|\dif x= \frac{1}{2\pi}\int_{-\infty}^{+\infty}e^{-\frac{(a+bx)^2}{2}}e^{-\frac{x^2}{2}}\dif x=\frac{1}{\sqrt{2\pi}\sqrt{1+b^2}}e^{-\frac{a^2}{2(1+b^2)}}<\infty,$$
	we have $\int_{-\infty}^{+\infty}\frac{\partial}{\partial a}g(a,x)\dif x$ convergent uniformly on $a\in\mathbb{R}$. Therefore, 
	\begin{equation}
	\begin{split}
	f'(a)&=\int_{-\infty}^{+\infty}\frac{\partial}{\partial a}g(a,x)\dif x\\
	&=\frac{1}{\sqrt{2\pi}\sqrt{1+b^2}}e^{-\frac{a^2}{2(1+b^2)}}.
	\end{split}
	\label{e fprime}
	\end{equation}
	And control convergence theorem gives 
	\begin{equation}
	\lim\limits_{a\rightarrow-\infty}f(a)=\mathbb{E}\lim\limits_{a\rightarrow-\infty}\Phi(a+b\xi)=0.
	\label{e f0}
	\end{equation}
	Combining (\ref{e defn f}), (\ref{e fprime}) and (\ref{e f0}) yields
	$$f(a)=\int_{-\infty}^{a}\frac{1}{\sqrt{2\pi}\sqrt{1+b^2}}e^{-\frac{x^2}{2(1+b^2)}}\dif x=\Phi\bigl(\frac{a}{\sqrt{1+b^2}}\bigr).$$
\end{proof}

In fact, substituting the above result and equation (\ref{e pi0(x)}) into equation (\ref{e prob with infer}) gives the same conclusion in equation (4.4) in \cite{myself}. The ``coincidence'' here inspires us that our computation can be greatly simplified by approximation through normal distribution. This is due to not only Proposition \ref{t exphi}, but also the fact that the sum of two independent and normally distributed random variables is also normally distributed. Now we derive the approximation expression of $p_{i,j}(x)$ in the $m$-$n$ arena with uniform fluctuations, where all individuals' coefficient of fluctuations is $\rho$.

\begin{thm}
	In an $m$-$n$ arena with uniform fluctuations, assume all players' coefficients of fluctuations are $\rho>0$. If $X_{i,0}\sim N(\mu,\sigma^2)$, then
	\begin{equation}
	\mathbb{E}X_{i+1,0}=\mu+\frac{2\sigma^2}{\sqrt{2\pi(2\sigma^2+\rho^2)}},
	\label{e_ex}
	\end{equation}
	\begin{equation}
	VarX_{i+1,0}=\sigma^2\left[1-\frac{2\sigma^2}{\pi(2\sigma^2+\rho^2)}\right].
	\label{e_varx}
	\end{equation}
	\label{t ncx}
\end{thm}

\begin{proof}
	Suppose $\xi\sim N(0,1)$, $p(\cdot)$ is its probability density. By Theorem \ref{t dens of str1} and Proposition \ref{t exphi} we have
	\[\begin{split}
	p_{i+1,0}(x)&=2p_{i,0}(x)\int_{-\infty}^{+\infty}\Phi\bigl(\frac{x-y}{\rho}\bigr)p_{i,0}(y)\dif y\\
	&=2p_{i,0}(x)\int_{-\infty}^{+\infty}\Phi\bigl(\frac{x-\mu-\sigma z}{\rho}\bigr)p(z)\dif z\\
	&=2p_{i,0}(x)\mathbb{E}\Phi\bigl(\frac{x-\mu}{\rho}-\frac{\sigma}{\rho}\xi\bigr)=2p_{i,0}(x)\Phi\bigl(\frac{x-\mu}{\sqrt{\sigma^2+\rho^2}}\bigr).
	\end{split}\]
	It follows that
	\[\begin{split}
	\mathbb{E}X_{i+1,0}&=2\int_{-\infty}^{+\infty}xp_{i,0}(x)\Phi\bigl(\frac{x-\mu}{\sqrt{\sigma^2+\rho^2}}\bigr)\dif x\\
	&=2\int_{-\infty}^{+\infty}(\mu+\sigma t)p(t)\Phi\bigl(\frac{\sigma}{\sqrt{\sigma^2+\rho^2}}t\bigr)\dif t=\mu+\frac{2\sigma^2}{\sqrt{2\pi(2\sigma^2+\rho^2)}}.
	\end{split}\]
	Further,
	\[\begin{split}
	\mathbb{E}X_{i+1,0}^2&=2\int_{-\infty}^{+\infty}x^2p_{i,0}(x)\Phi\bigl(\frac{x-\mu}{\sqrt{\sigma^2+\rho^2}}\bigr)\dif x\\
	&=2\int_{-\infty}^{+\infty}(\mu+\sigma t)^2p(t)\Phi\bigl(\frac{\sigma}{\sqrt{\sigma^2+\rho^2}}t\bigr)\dif t\\
	&=\mu^2+\frac{4\mu\sigma^2}{\sqrt{2\pi(2\sigma^2+\rho^2)}}+\sigma^2.
	\end{split}\]
	Therefore,
	$$VarX_{i+1,0}=\mathbb{E}X_{i+1,0}^2-(\mathbb{E}X_{i+1,0})^2=\sigma^2\left[1-\frac{2\sigma^2}{\pi(2\sigma^2+\rho^2)}\right].$$
\end{proof}

Similarly, we can derive the recursion equations of the expectation and variance of $X_{i,j}$, provided that $X_{i,j}$ is approximately normally distributed. 

\begin{thm}
	In an $m$-$n$ arena with uniform fluctuations, assume all players' coefficients of fluctuations are $\rho>0$. Then $X_{i,j}\sim N(\mu_{i,j},\sigma_{i,j}^2)\ (0\leqslant i\leqslant m,0\leqslant j\leqslant n)$ holds approximately, where $(\mu_{i,j},\sigma_{i,j}^2)$ satisfy
	
	\begin{equation}
	\left\{
	\begin{split}
	\sigma_{0,0}^2&=1,\\
	\sigma_{i,0}^2&=\sigma_{i-1,0}^2\left[1-\frac{2\sigma_{i-1,0}^2}{\pi(2\sigma_{i-1,0}^2+\rho^2)}\right],\ (1\leqslant i\leqslant m-1)\\
	\sigma_{0,j}^2&=\sigma_{0,j-1}^2\left[1-\frac{2\sigma_{0,j-1}^2}{\pi(2\sigma_{0,j-1}^2+\rho^2)}\right],\ (1\leqslant j\leqslant n-1)\\
	\sigma_{i,j}^2&=\frac{i}{i+j}\sigma_{i-1,j}^2\left[1-\frac{2\sigma_{i-1,j}^2}{\pi(2\sigma_{i-1,j}^2+\rho^2)}\right]+\frac{j}{i+j}\sigma_{i,j-1}^2\left[1-\frac{2\sigma_{i,j-1}^2}{\pi(2\sigma_{i,j-1}^2+\rho^2)}\right],\\
	&\quad\quad(1\leqslant i\leqslant m-1,1\leqslant j\leqslant n-1)\\
	\sigma_{m,j}^2&=\sigma_{m-1,j}^2\left[1-\frac{2\sigma_{m-1,j}^2}{\pi(2\sigma_{m-1,j}^2+\rho^2)}\right],\ (0\leqslant j\leqslant n-1)\\
	\sigma_{i,n}^2&=\sigma_{i,n-1}^2\left[1-\frac{2\sigma_{i,n-1}^2}{\pi(2\sigma_{i,n-1}^2+\rho^2)}\right],\ (0\leqslant i\leqslant m-1)
	\end{split}\right.
	\label{e recur of mu}	
	\end{equation}
	and
	\begin{equation}
	\left\{
	\begin{split}
	\mu_{0,0}&=0,\\
	\mu_{i,0}&=\mu_{i-1,0}+\frac{2\sigma_{i-1,0}^2}{\sqrt{2\pi(2\sigma_{i-1,0}^2+\rho^2)}},\ (1\leqslant i\leqslant m-1)\\
	\mu_{0,j}&=\mu_{0,j-1}-\frac{2\sigma_{0,j-1}^2}{\sqrt{2\pi(2\sigma_{0,j-1}^2+\rho^2)}},\ (1\leqslant j\leqslant n-1)\\
	\mu_{i,j}&=\frac{i}{i+j}\left[\mu_{i-1,j}+\frac{2\sigma_{i-1,j}^2}{\sqrt{2\pi(2\sigma_{i-1,j}^2+\rho^2)}}\right]+\frac{j}{i+j}\left[\mu_{i,j-1}-\frac{2\sigma_{i,j-1}^2}{\sqrt{2\pi(2\sigma_{i,j-1}^2+\rho^2)}}\right],\\
	&\quad\quad(1\leqslant i\leqslant m-1,1\leqslant j\leqslant n-1)\\
	\mu_{m,j}&=\mu_{m-1,j}+\frac{2\sigma_{m-1,j}^2}{\sqrt{2\pi(2\sigma_{m-1,j}^2+\rho^2)}},\ (0\leqslant j\leqslant n-1)\\
	\mu_{i,n}&=\mu_{i,n-1}-\frac{2\sigma_{i,n-1}^2}{\sqrt{2\pi(2\sigma_{i,n-1}^2+\rho^2)}},\ (0\leqslant i\leqslant m-1)
	\end{split}\right.
	\label{e recur of mu}
	\end{equation}
	
	\label{t recur}
\end{thm}

We now derive an MAP estimator via the above approximation in the next section.

\section{Estimation of strengths and coefficients of fluctuations}

If we already know the past performance of a player in an $m$-$n$ arena with fluctuations and suppose he reaches $(i,j)\in S_{m,n}$ for $N_{i,j}$ times, where 
$$S_{m,n}=\bigl\{(m,j):j=0,1,\cdots,n-1\bigr\} \cup \bigl\{(i,n):i=0,1,\cdots,m-1\bigr\}.$$
Then the likelihood of these samples is 
\begin{equation}
P(x,\rho)=\prod_{(i,j)\in S_{m,n}}\mathbb{P}(\mathcal{A}_{i,j}|X_l=x,\rho_l=\rho)p_{0,0}(x)\stackrel{\text{approx.}}\propto\frac{\prod_{(i,j)\in S_{m,n}}\widetilde{p_{i,j}}^{N_{i,j}}(x)}{p_{0,0}^{N-1}(x)},
\label{e likelihood}
\end{equation}
where $\widetilde{p_{i,j}}(\cdot)$ is the PDF of Gaussian random variables with mean $\mu_{i,j}$ and variance $\sigma_{i,j}^2$, $N$ is the sum of $N_{i,j}$ on $S_{m,n}$.

In fact, this approach both makes no sense theoretically, and performs badly in practice. Based on our assumptions of arena models with fluctuations, Theorem \ref{t prob with infer2} gives a correct approach to compute the probability for one to obtain different final results, given his strength and coefficient of fluctuations, but equation (\ref{e likelihood}) is using a bad approximation of those probability density involved in. A normal approximation is a doable simplification for computing the distribution function, but not a satisfying way to approximate the density function. Besides, the probabilities no longer sum up to 1 for different final results, if we approximately compute by
\begin{equation}
    \mathbb{P}(\mathcal{A}_{i,j}|X_l=x,\rho_l=\rho)\approx\binom{i+j-1}{i-1}(\frac{1}{2})^{i+j}\cdot\frac{\widetilde{p_{i,j}}(x)}{p_{0,0}(x)}.
\end{equation}
The simulation results also show that this approach has a poor estimation on the coefficient of fluctuations, which drives us to think about another estimation method.

Return to the basic assumption of an arena with uniform fluctuations (suppose the coefficient of fluctuations of the arena is $\rho$). The event that a player who is in the state $(i,j)$ currently with strength $x$ will get into the state $(i+1,j)$, is equivalent to the random event that $Y_1=x+\frac{\rho}{\sqrt{2}}\epsilon_1>Y_2=X_{i,j}+\frac{\rho}{\sqrt{2}}\epsilon_2$. Here $Y_1$ and $Y_2$ are respectively the performance of this player and his next opponent, $\epsilon_1$ and $\epsilon_2$ are respectively the relative fluctuations of this player and his opponent. We use a random variable $X_{i,j}$ to describe the strength of his next opponent, which is approximated by a Gaussian distribution with mean $\mu_{i,j}$ and variance $\sigma_{i,j}^2$ in theorem \ref{t recur}. Therefore, that conditional probability can be easily given by $\Phi\left((x-\mu_{i,j})/\sqrt{\sigma_{i,j}^2+\rho^2}\right)$. Then we can compute other conditional probability similarly and derive the probabilities of a player with strength $x$ to obtain different final results step by step. For instance, in a 2-2 arena with uniform fluctuations (suppose the coefficient of fluctuations is $\rho$), we have the following approximation: 
\begin{equation}\label{eq:prob}
    \left\{\begin{split}
        \mathbb{P}_{\rho}(\mathcal{A}_{2,0}|X_l=x)&\approx\Phi(\frac{x}{\sqrt{1+\rho^2}})\Phi(\frac{x-\mu_{1,0}}{\sqrt{\sigma_{1,0}^2+\rho^2}})\\
        \mathbb{P}_{\rho}(\mathcal{A}_{2,1}|X_l=x)&\approx\Bigl[\Phi(\frac{x}{\sqrt{1+\rho^2}})\Phi(\frac{\mu_{1,0}-x}{\sqrt{\sigma_{1,0}^2+\rho^2}})+\Phi(\frac{-x}{\sqrt{1+\rho^2}})\Phi(\frac{x-\mu_{0,1}}{\sqrt{\sigma_{0,1}^2+\rho^2}})\Bigr]\Phi(\frac{x-\mu_{1,1}}{\sqrt{\sigma_{1,1}^2+\rho^2}})\\
        \mathbb{P}_{\rho}(\mathcal{A}_{1,2}|X_l=x)&\approx\Bigl[\Phi(\frac{x}{\sqrt{1+\rho^2}})\Phi(\frac{\mu_{1,0}-x}{\sqrt{\sigma_{1,0}^2+\rho^2}})+\Phi(\frac{-x}{\sqrt{1+\rho^2}})\Phi(\frac{x-\mu_{0,1}}{\sqrt{\sigma_{0,1}^2+\rho^2}})\Bigr]\Phi(\frac{\mu_{1,1}-x}{\sqrt{\sigma_{1,1}^2+\rho^2}})\\
        \mathbb{P}_{\rho}(\mathcal{A}_{0,2}|X_l=x)&\approx\Phi(\frac{-x}{\sqrt{1+\rho^2}})\Phi(\frac{\mu_{0,1}-x}{\sqrt{\sigma_{0,1}^2+\rho^2}})
        \end{split}\right.
\end{equation}

Through this approximation, we can derive a new approximation of the product of likelihood and prior probability by
\begin{equation}
    P(x,\rho)\approx\prod_{(i,j)\in S}\mathbb{P}_{\rho}(\mathcal{A}_{i,j}|X_l=x)p_{0,0}(x)
    \label{e likelihood2}
\end{equation}
to compute our MAP estimator of $(x,\rho)$. Notice that we have not already proved the effectiveness of this approximation on distributions, but we will show its good performance in practice in Section 6.

\section{Prediction of future results}

In practice, we want to not only rank players by estimating their strengths, but also predict their future performance from their past results, which is of great significance in sports, psychology (stimulus are strengths and physiological reflections are results), and species competitions. In this section we will briefly discuss two prediction approaches in arena model with fluctuations.

Since Section 4 provides an easy way to estimate players' strengths and coefficient of fluctuations under the rule of MAP, we can directly substitute the results into the conditional probability given one's strength and coefficient of fluctuations (for instance, equation (\ref{eq:prob}) when $m=n=2$). Besides, we can also apply Bayesian posterior distribution into the prediction by integrating the strength $x$ (we recommend to treat $\rho$ as a constant rather than a random variable to avoid expensive computation). In the rest of the paper, we only use the first method to implement simulations and applications in FIFA World Cup, even though both are feasible and effective.

\section{Tests and applications}

In this section, we first conduct a simulation test for the estimators addressed in Section 4 and use those estimates to predict the future results of individuals following Section 5. Then we apply our estimation method to the real sample data from FIFA Word Cup.

\subsection{Simulations}

In this part, we test the performance of our estimator in a 2-2 arena with uniform fluctuations. As for the power of prediction, we compare the result from our approximation method with a classic method which uses empirical frequency to estimate the real probability of a player ending with a specific result.

\subsubsection{Estimation of strengths and coefficients of fluctuations}

This part shows the estimation of a player's strength and coefficient of fluctuations by equation (\ref{eq:prob}) and (\ref{e likelihood2}). Specifically, we assume all players have the same coefficients of fluctuations denoted by $\rho$. For $\rho=0.1,0.5,1.0$ respectively, we study the player A with his strength $X_1$ equal to $0,0.01,0.02,\cdots,2.00$ (we conduct simulation on 201 discrete points). It should emphasized that the ``player A" or``player B" does not refer to their strength or final results. The order is randomly given and never changes in the simulation part. Due to the symmetry of the 2-2 arena, we leave out the implementations for cases that $X_1=-0.01,-0.02,\cdots,-2.00$. Let the player plays in a 2-2 arena for $r=20$ and 80 times respectively with 1023 competitors, whose strengths are sampled independently from the standard normal distribution.

Intuitively, the final results of a player in one arena run tends to have larger fluctuations as $\rho$ increases, which affects the estimation a lot. The randomness of samples is derived from both random matches with other individuals and fluctuations in each round. Therefore, for the case that $\rho=1$, we set $r=80$ to mitigate the influence of randomness from matching towards the estimation of $\rho$ and $X_1$.

\begin{figure}[!h]
\subfigure[Estimation of $\rho$]{
\begin{minipage}[t]{0.5\linewidth}
\centering
\includegraphics[width=8cm,height=4cm]{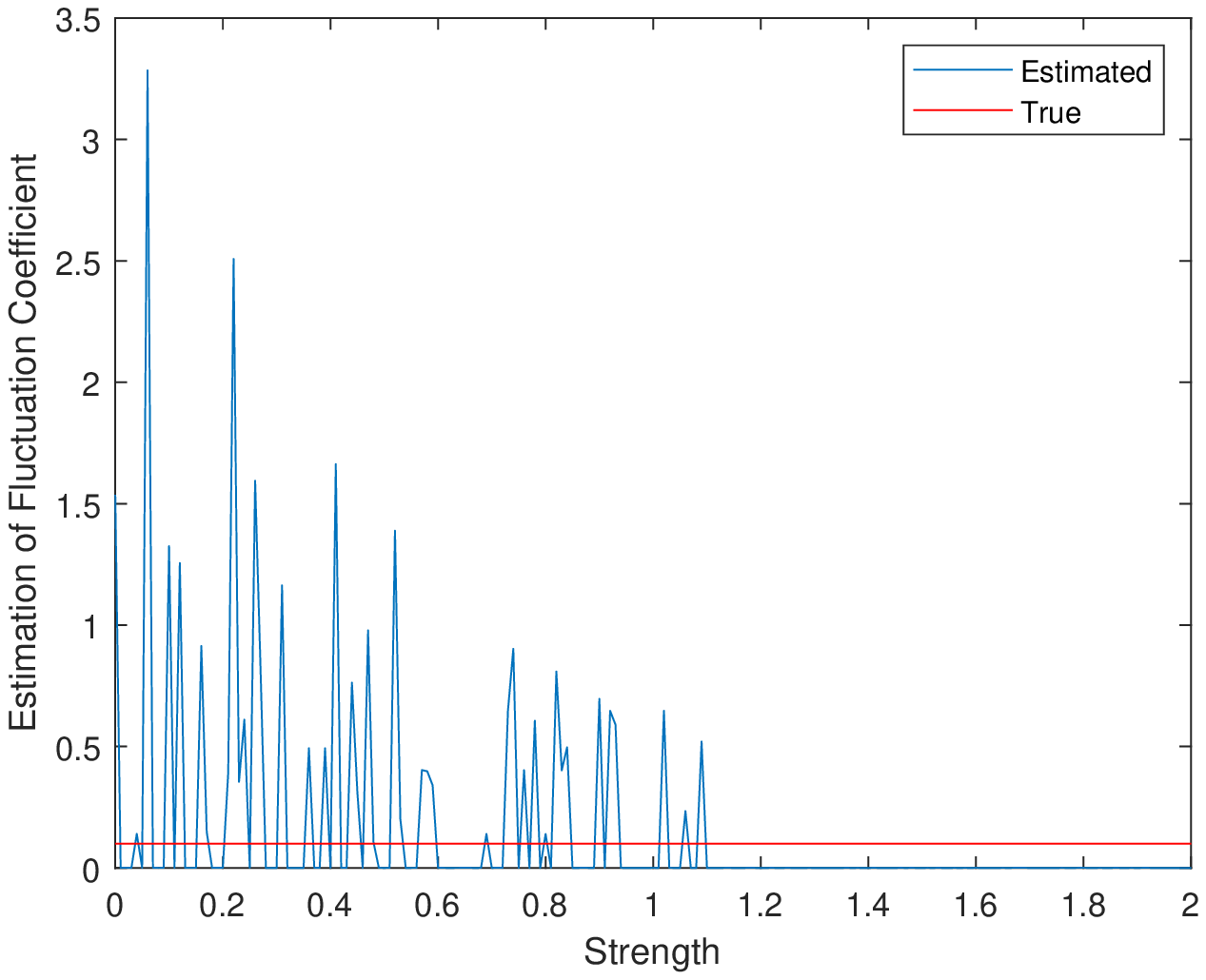}
\end{minipage}
}
\subfigure[Estimation of $X_1$]{
\begin{minipage}[t]{0.5\linewidth}
\centering
\includegraphics[width=8cm,height=4cm]{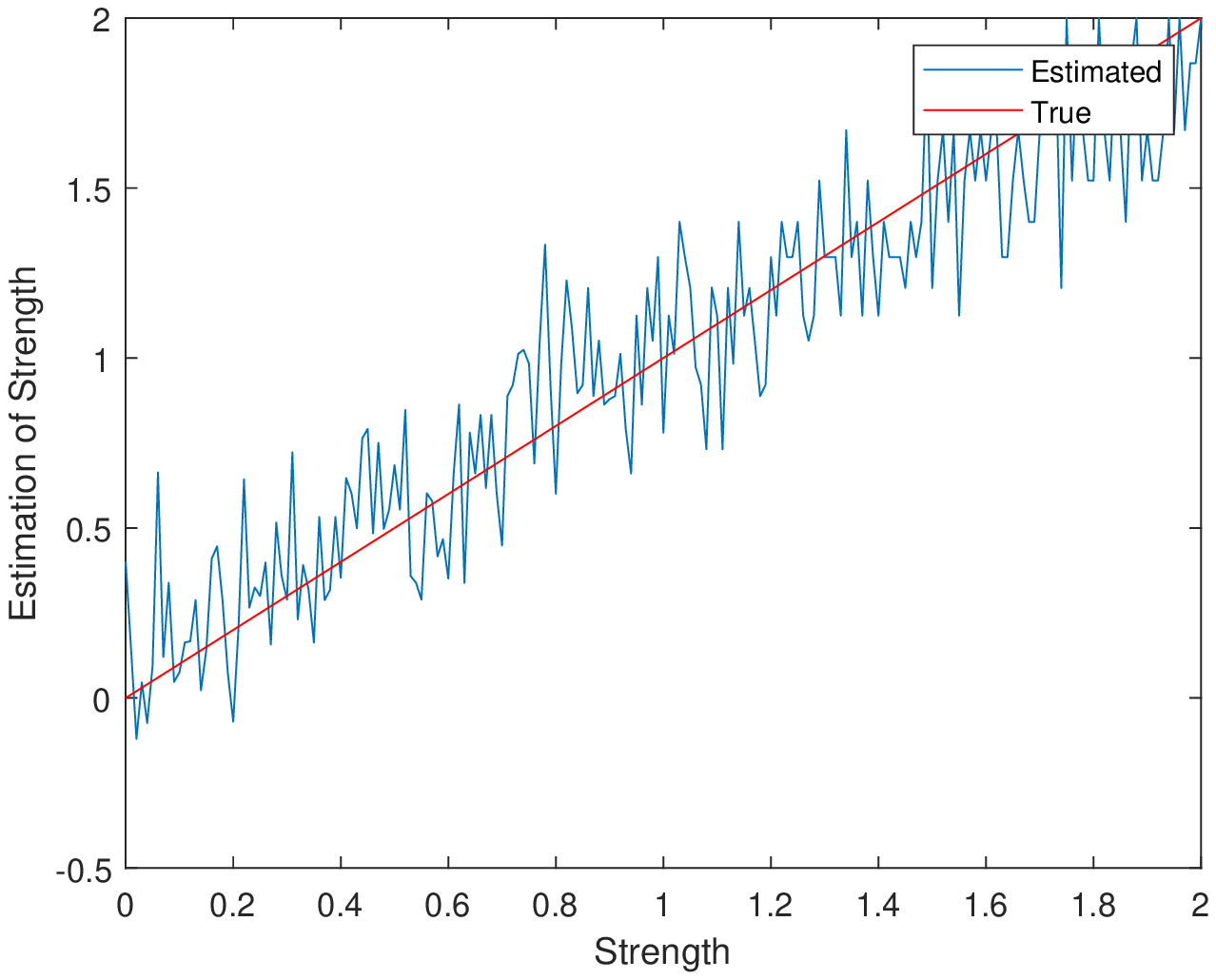}
\end{minipage}
}
\caption{The Figures 2a and 2b illustrate the estimation results of player A's coefficient of fluctuations $\hat{\rho}$ and strength $\hat{X_1}$ respectively when $\rho=0.1$ and $r=20$, as $X_1$ increases from 0 to 2.00.}
\end{figure}

\begin{figure}[!h]
\subfigure[Estimation of $\rho$]{
\begin{minipage}[t]{0.5\linewidth}
\centering
\includegraphics[width=8cm,height=4cm]{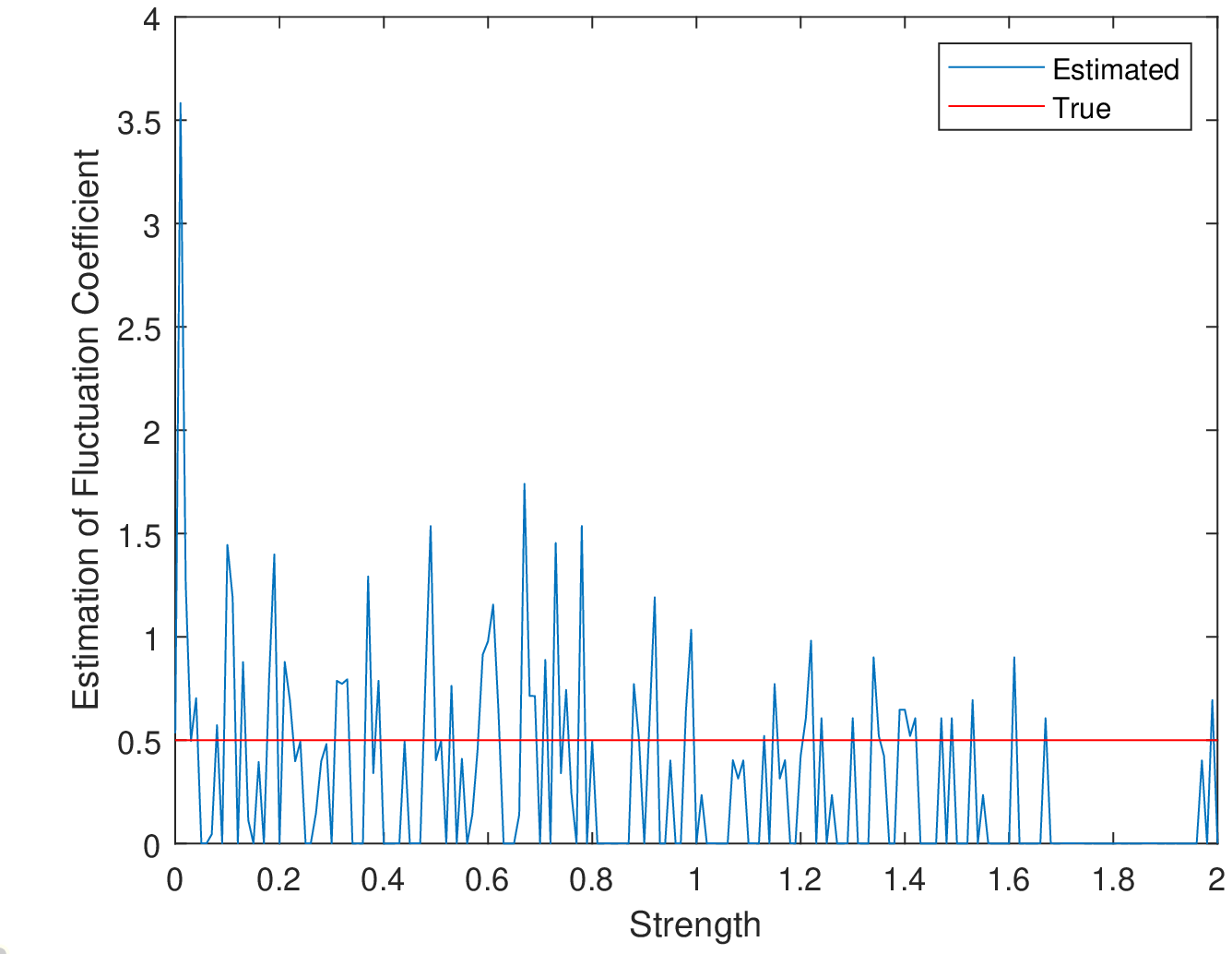}
\end{minipage}
}
\subfigure[Estimation of $X_1$]{
\begin{minipage}[t]{0.5\linewidth}
\centering
\includegraphics[width=8cm,height=4cm]{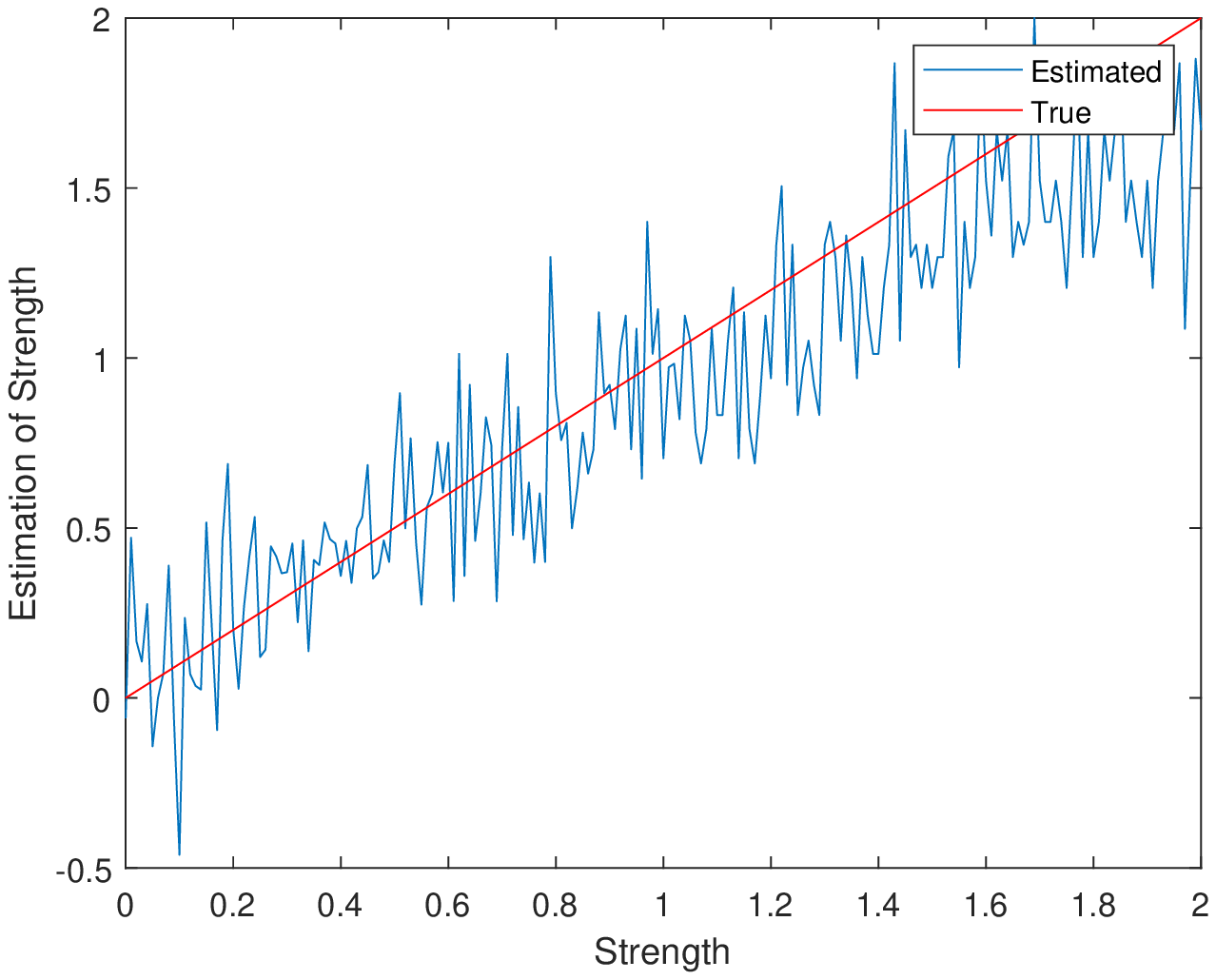}
\end{minipage}
}
\caption{The Figures 3a and 3b illustrate the estimation results of player A's coefficient of fluctuations $\hat{\rho}$ and strength $\hat{X_1}$ respectively when $\rho=0.5$ and $r=20$, as $X_1$ increases from 0 to 2.00.}
\end{figure}

\begin{figure}[!h]
\subfigure[Estimation of $\rho$]{
\begin{minipage}[t]{0.5\linewidth}
\centering
\includegraphics[width=8cm,height=4cm]{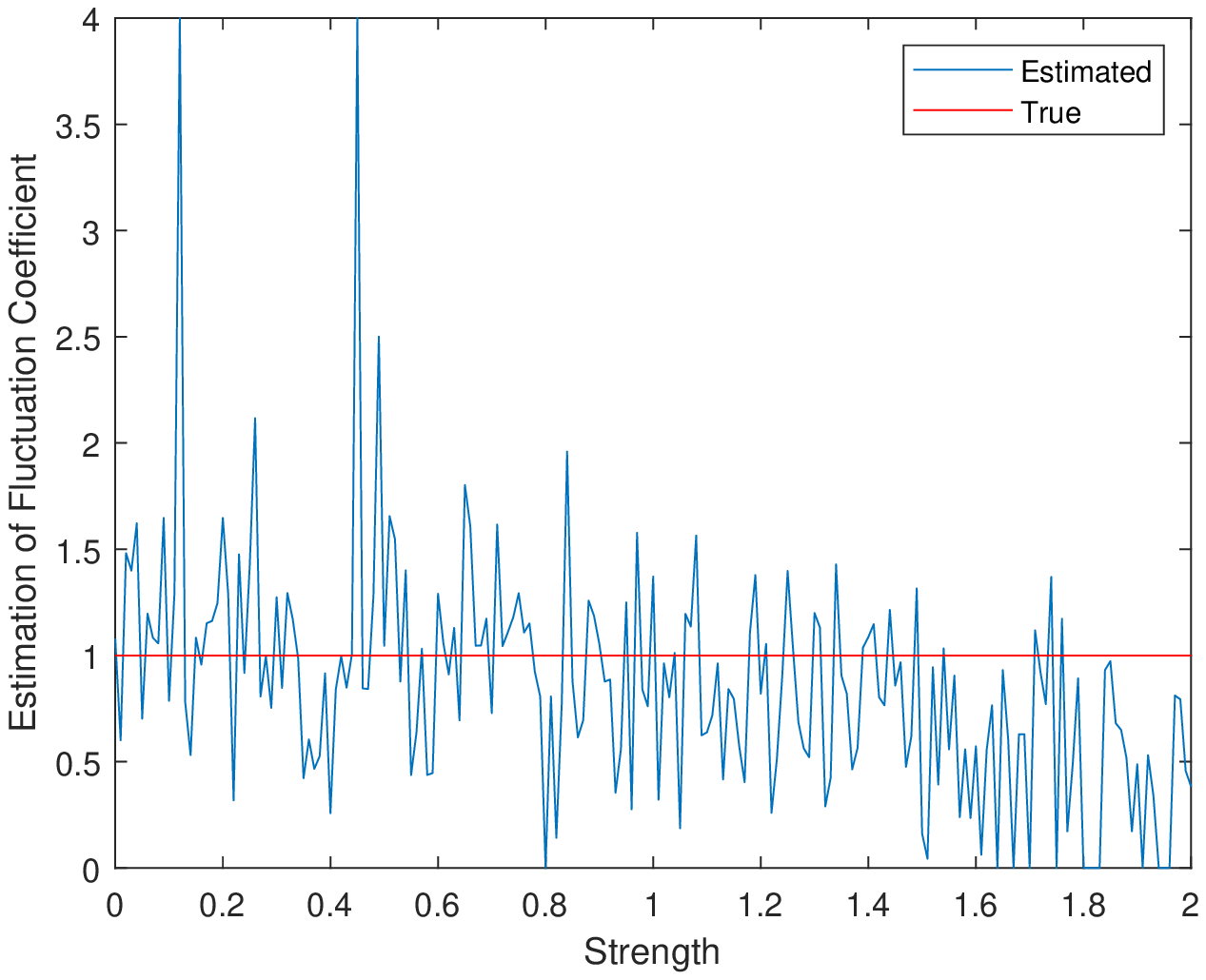}
\end{minipage}
}
\subfigure[Estimation of $X_1$]{
\begin{minipage}[t]{0.5\linewidth}
\centering
\includegraphics[width=8cm,height=4cm]{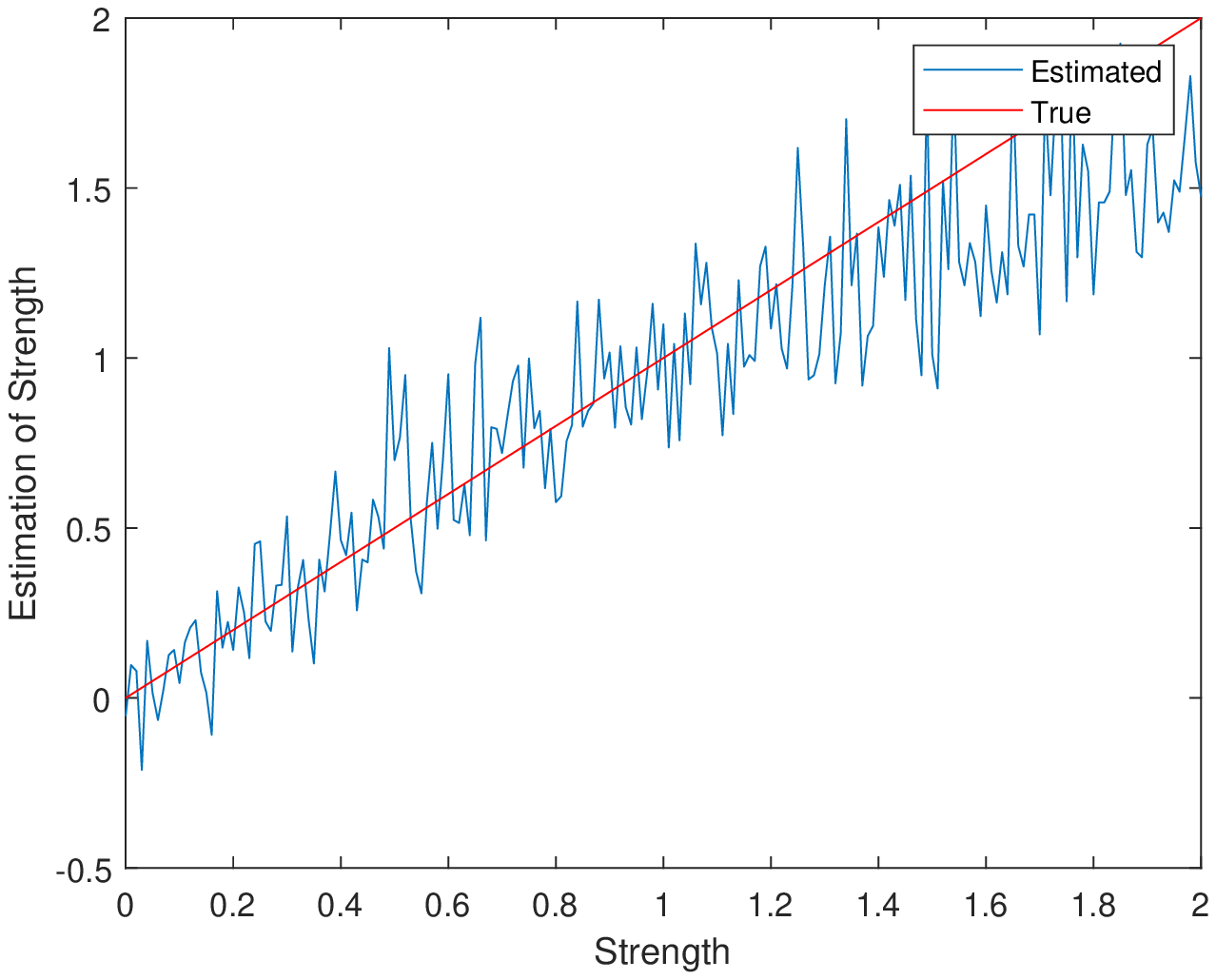}
\end{minipage}
}
\caption{The Figures 4a and 4b illustrate the estimation results of player A's coefficient of fluctuations $\hat{\rho}$ and strength $\hat{X_1}$ respectively when $\rho=1.0$ and $r=80$, as $X_1$ increases from 0 to 2.00.}
\end{figure}

It could be found in the Figure 2a that when $rho$ is extremely low (such as $\rho=0.1$), the estimation performs badly for small $X$. This is predictable since the randomness from random matching denominates. We can study $\beta=\frac{1}{1+\rho}$ as an transformation of $\rho$ to decrease absolute error. Besides, as for an individual whose strength is medium (fairly close to zero), the estimation of his coefficient of fluctuations is inevitably much larger than the true value. In this case the estimate is greatly sensitive to ``exceptional'' results, which also shows up frequently due to the random match. One solution to this is to estimate with more data since the sample size as large as 20 is not easy to compensate the randomness from pairing. And the other one is to increase the $m$ and $n$ to make the final results more discriminating.

\subsubsection{Prediction of future results}
In this section, we present the estimated probabilities that player A ends with results $(2,0),(2,1),(1,2),(0,2)$ in an arena run, denoted by $p_{20},p_{21},p_{12},p_{02}$ respectively. We also compare our method with a direct method which uses empirical frequencies to estimate real probabilities. Specifically, consider a 2-2 arena consisting of 1024 players with coefficient of fluctuations $\rho=0.5$. Assume that player A with strength $X_1=0,0.01,0.02,\cdots,2$ plays with the other 1023 competitors for $r=20$ times. With these results, we have two estimates of $p_{20},p_{21},p_{12},p_{02}$ respectively by our approximation method in arena model and the frequency method.

\begin{figure}[!h]
\subfigure[Estimation of $p_{20}$]{
\begin{minipage}[t]{0.5\linewidth}
\centering
\includegraphics[width=8cm,height=4cm]{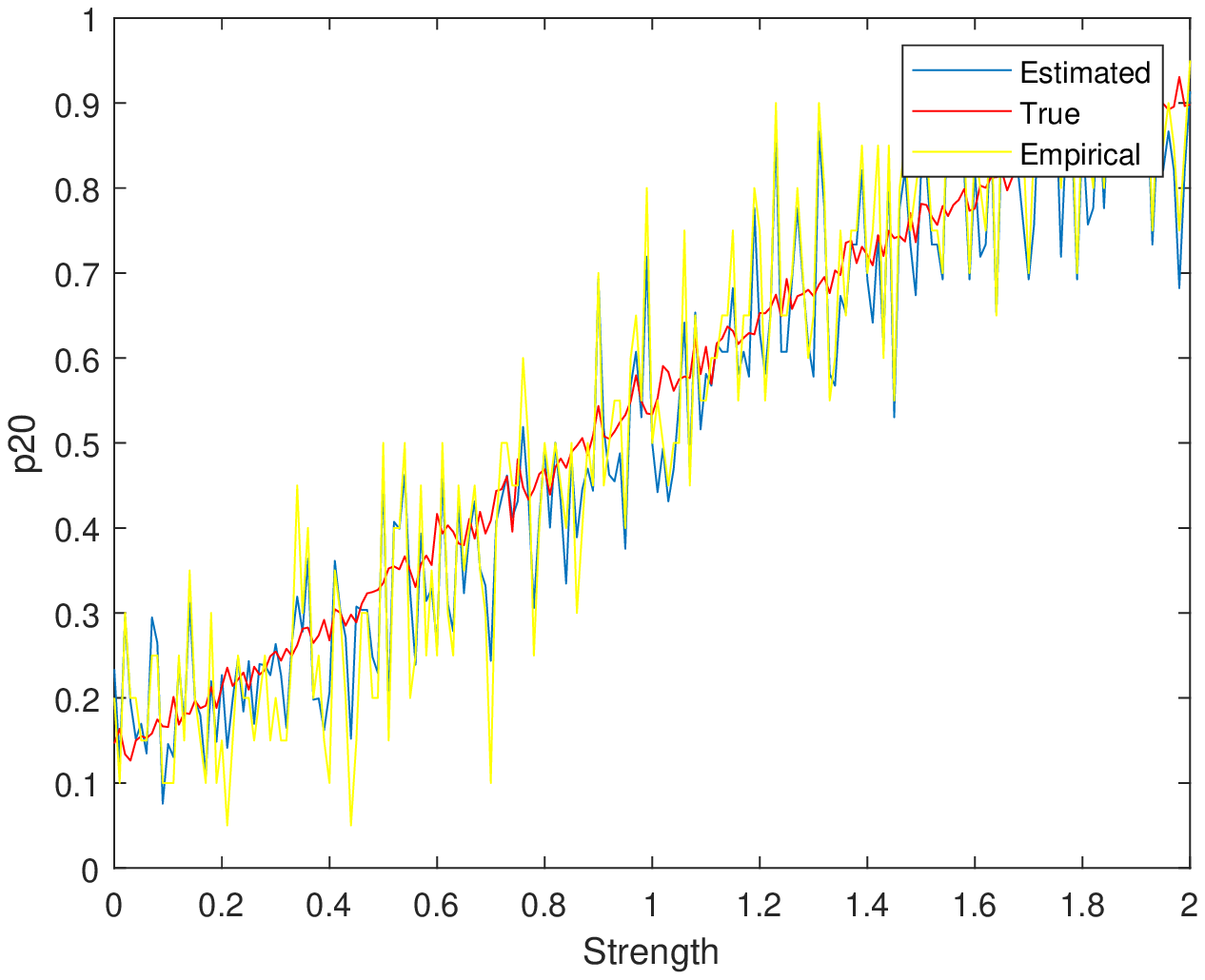}
\end{minipage}
}
\subfigure[Estimation of $p_{21}$]{
\begin{minipage}[t]{0.5\linewidth}
\centering
\includegraphics[width=8cm,height=4cm]{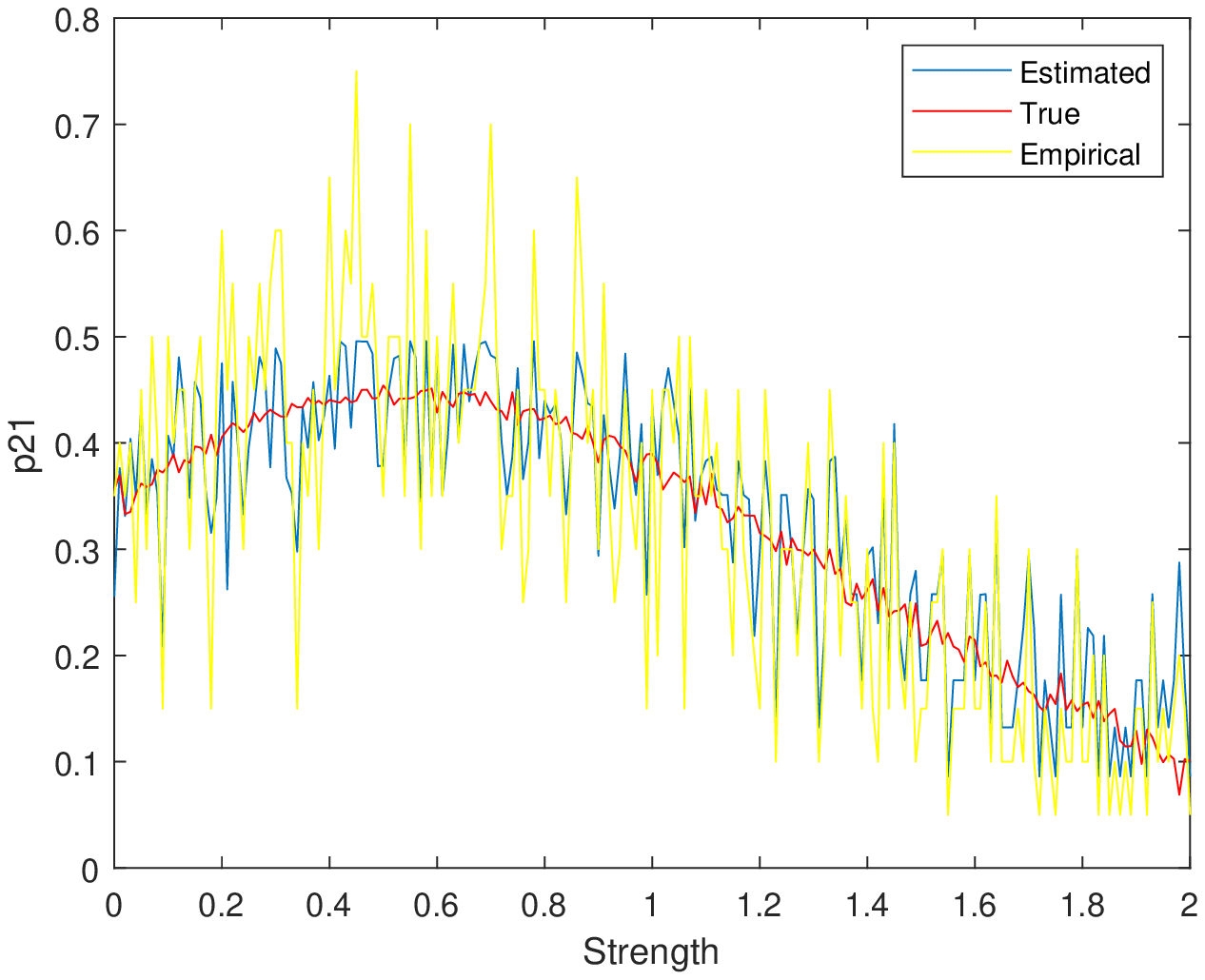}
\end{minipage}
}
\end{figure}

\begin{figure}[!h]
\subfigure[Estimation of $p_{12}$]{
\begin{minipage}[t]{0.5\linewidth}
\centering
\includegraphics[width=8cm,height=4cm]{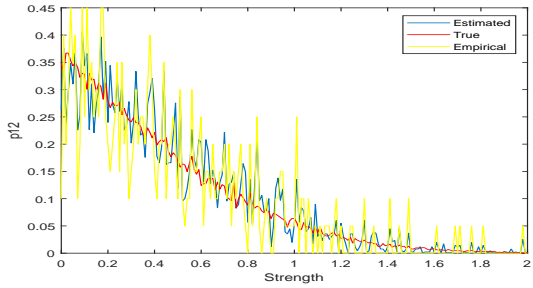}
\end{minipage}
}
\subfigure[Estimation of $p_{02}$]{
\begin{minipage}[t]{0.5\linewidth}
\centering
\includegraphics[width=8cm,height=4cm]{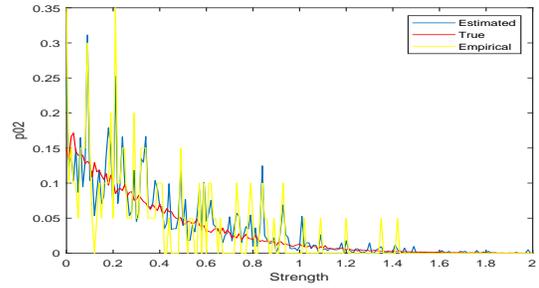}
\end{minipage}
}
\caption{The Figures 5a-5d present the estimates of $p_{20},p_{21},p_{12},p_{02} $ through the approximation method in arena model and the frequency approach when $\rho=0.5$ and $r=20$.}
\end{figure}

Even though the estimation of strengths flip around the true value as Figure 3b presents, arena model shows astounding advantages over the nonparametric frequency estimation when applied to predict future results. Firstly, our estimator has much smaller volatility thus is more stable than estimating by frequencies, especially when the sample size is relatively small. More importantly, the estimation of, for example, $p_{20}$ may equal to zero by frequency approach, since there is possibility that player A have never reached $(2,0)$ within several runs, especially when $r$ is small. We do not have such problem when apply the method of this paper.

\subsection{Application in the FIFA World Cup}

Now we apply our estimation method to some real data from FIFA World Cup. In each FIFA World Cup, teams entering top 16 will participate in the knockout which is modelled as a 5-1 arena with fluctuations. We collect 20 final results of four countries, in which `0' means the team did not enter top 16, and `1', `2', `3', `4', `5' present entering top 16, 8, 4, 2, 1 respectively. All data we use in this part are available at \cite{fifa}. We first use the ten final results in Table \ref{table1} to estimate the strength and coefficient of fluctuations for every country, and use those estimates to predict the probability for every country to reach every final result. We derive the real probabilities from Table \ref{table2} and compare in Table \ref{table3} the predictions by our method with those by frequencies.

\begin{table}[!htb]
\centering
\caption{Data for train}\label{table1}
\begin{tabular}{ccccccccccc}
\toprule
Country& 1930& 1938& 1954& 1962& 1970& 1978& 1986& 1994& 2002& 2010\\
\midrule
Brazil& 3& 3& 2& 5& 5& 3& 2& 5& 5& 2\\
Italy& 0& 5& 1& 1& 4& 3& 1& 4& 1& 0\\
Argentina& 4& 0& 0& 1& 0& 5& 5& 1& 0& 2\\
Sweden& 0& 3& 0& 0& 1& 1& 0& 3& 1& 0\\
\bottomrule
\end{tabular}
\end{table}

\begin{table}[!htb]
\centering
\caption{Data for test}\label{table2}
\begin{tabular}{ccccccccccc}
\toprule
Country& 1934& 1950& 1958& 1966& 1974& 1982& 1990& 1998& 2006& 2014\\
\midrule
Brazil& 1& 4& 5& 1& 3& 2& 1& 4& 2& 3\\
Italy& 5& 2& 0& 1& 1& 5& 3& 2& 5& 0\\
Argentina& 1& 0& 1& 2& 2& 1& 4& 2& 2& 4\\
Sweden& 2& 3& 4& 0& 2& 0& 0& 0& 1& 0\\
\bottomrule
\end{tabular}
\end{table}

Table \ref{table3} presents our estimates of strength and coefficient of fluctuations for each country. With these estimates, we can predict the probability for every country to reach every final result. Column P1 contains the predictions by our method and P2 denotes the predictions by frequencies derived from Table \ref{table1}. If we approximately treat the frequencies in Table \ref{table2} as real probabilities which is displayed in column F, we can compare the two predictions P1 and P2 by their Euclidean distances to the ``real'' probabilities F. 

\begin{table}[!h]
\centering
\begin{tabular}{|c|c|c|c|c|c|c|c|c|c|c|c|c|}
\hline
\multirow{2}*{$\xi$}& \multicolumn{3}{|c|}{Brazil}& \multicolumn{3}{|c|}{Italy}& \multicolumn{3}{|c|}{Argentina}& \multicolumn{3}{|c|}{Sweden}\\
\cline{2-13}
& \multicolumn{3}{|c|}{$X=1.80,\rho=0.50$}& \multicolumn{3}{|c|}{$X=1.17,\rho=1.56$}& \multicolumn{3}{|c|}{$X=1.14,\rho=3.32$}& \multicolumn{3}{|c|}{$X=0.06,\rho=2.11$}\\
\cline{2-13}
& F& P1& P2& F& P1& P2& F& P1& P2& F& P1& P2\\
\hline
0& 0& 0.05& 0& 0.2& 0.26& 0.2& 0.1& 0.37& 0.4& 0.4& 0.49& 0.5\\
\hline
1& 0.3& 0.09& 0& 0.2& 0.24& 0.4& 0.3& 0.25& 0.2& 0.2& 0.28& 0.3\\
\hline
2& 0.2& 0.15& 0.3& 0.2& 0.20& 0& 0.4& 0.16& 0.1& 0.2& 0.14& 0\\
\hline
3& 0.2& 0.20& 0.3& 0.1& 0.14& 0.1& 0& 0.10& 0& 0.1& 0.06& 0.2\\
\hline
4& 0.2& 0.21& 0& 0& 0.08& 0.2& 0.2& 0.06& 0.1& 0.1& 0.02& 0\\
\hline
5& 0.1& 0.30& 0.4& 0.3& 0.07& 0.1& 0& 0.06 & 0.2& 0& 0.01& 0\\
\hline
\end{tabular}
\caption{Results within different methods}\label{table3}
\end{table}

The comparisons in Table \ref{table4} indicates that our method achieves better predictions than simply predicting by frequencies in the sense of Euclidean distance error. The $d(\cdot,\cdot)$ here indicates the Euclidean distance of two sets of probabilities for classification.

\begin{table}[!htbp]
\centering
\begin{tabular}{|c|c|c|c|c|}
\hline
& Brazil& Italy& Argentina& Sweden\\
\hline
$d(P1,F)$& 0.30& 0.26& 0.41& 0.16\\
\hline
$d(P2,F)$& 0.49& 0.40& 0.49& 0.28\\
\hline
\end{tabular}
\caption{The comparison between prediction results P1 and P2}\label{table4}
\end{table}

Table 5 presents the estimates and predictions using all of the twenty final results in Table \ref{table1} and \ref{table2}. We present the estimate of strengths of these four teams and the results match our expectation that Brazil team is ``stonger" than Italy team, Italy team is ``stronger" than Argentina team and so on. It shows another advantage of our method over predicting by frequencies. We can also see that even we have 20 past results of a country, the probability such as $p_{01}$ or $p_{50}$ is likely to be estimated to zero by frequency, while we do not have this problem using our method.

\begin{table}[!htbp]
	\centering
	\begin{tabular}{|c|c|c|c|c|c|c|c|c|}
		\hline
		\multirow{3}*{$\xi$}& \multicolumn{2}{|c|}{Brazil}& \multicolumn{2}{|c|}{Italy}& \multicolumn{2}{|c|}{Argentina}& \multicolumn{2}{|c|}{Sweden}\\
		\cline{2-9}
		& \multicolumn{2}{|c|}{$X=1.64,\rho=0.55$}& \multicolumn{2}{|c|}{$X=1.56,\rho=1.87$}& \multicolumn{2}{|c|}{$X=1.18,\rho=1.62$}& \multicolumn{2}{|c|}{$X=0.26,\rho=2.46$}\\
		\cline{2-9}
		& F& P& F& P& F& P& F& P\\
		\hline
		0& 0& 0.07& 0.2& 0.23& 0.25& 0.28& 0.5& 0.46\\
		\hline
		1& 0.15& 0.13& 0.3& 0.22& 0.25& 0.24& 0.2& 0.27\\
		\hline
		2& 0.25& 0.18& 0.1& 0.18& 0.25& 0.19& 0.1& 0.15\\
		\hline
		3& 0.25& 0.22& 0.1& 0.14& 0& 0.13& 0.15& 0.07\\
		\hline
		4& 0.1& 0.20& 0.1& 0.10& 0.15& 0.08& 0.05& 0.03\\
		\hline
		5& 0.25& 0.20& 0.2& 0.13& 0.1& 0.08& 0& 0.02\\
		\hline
	\end{tabular}
	\caption{Results within different methods}
	\label{table5}
\end{table}

\section{Conclusion}

First, we propose an efficient estimation method for arena with fluctuations in this paper, which gains satisfying results in both simulations and applications. The approximation gets involved in the arena model in mainly two parts: the assumption that all individuals have the same coefficient of fluctuations and the use of normal distribution function as the CDF of players' strengths in different states. It can be further improved by appropriately adjusting the moments in Theorem \ref{t recur} through simulations. It is also a problem worthy of study that how to optimize estimation results after obtaining raw estimates. A recursion algorithm for estimates maybe helpful to obtain optimal results.

In addition, Zhang proves a significant property of arenas without fluctuations in \cite{myself} that the prediction results are invariant provided that choosing an appropriate prior. However, this property is no longer maintained in arenas with fluctuations, that is, the prediction results are related with the probability distribution assumed by us. In this paper, we do not study the difference in prediction results when using various distributions and leave it to further research.

\section*{\centering{Acknowledgements}}

We are grateful to Prof. W. Huang and Prof. Q.-H. Zhang for many useful discussions and suggestsions on systemizing our ideas and polishing this thesis.

\end{document}